\documentstyle[pra,aps,multicol]{revtex}
\begin{document}

\title{$\Lambda$'s, V's, and optimal cloning with stimulated emission}

\author{ J. Kempe$^1$
  \thanks{also \'Ecole Nationale Superieure des
    T\'el\'ecommunications, Paris, France}, C. Simon$^2$ , and G. Weihs$^2$\thanks{Emails:
    kempe@math.berkeley.edu,
    Christoph.Simon@univie.ac.at, Gregor.Weihs@univie.ac.at}}

\address{$^1$ Departments of Mathematics and Chemistry$^{}$\\ University  of California, Berkeley\\ $^2$ Institut f\"ur Experimentalphysik, Universit\"at Wien, Boltzmanngasse 5,\\ A-1090 Wien,  Austria\\ }

\date{\today}

\maketitle

\begin{abstract}
We show that optimal universal cloning of the polarization state
of photons can be achieved via stimulated emission in three-level
systems, both of the Lambda and the V type. We establish the
equivalence of our systems with coupled harmonic oscillators,
which permits us to analyze the structure of the cloning
transformations realized. These transformations are shown to be
equivalent to the optimal cloning transformations for qubits
discovered by Bu\v{z}ek and Hillery, and Gisin and Massar. The
down-conversion cloner discovered previously by some of the
authors is obtained as a limiting case. We demonstrate an
interesting equivalence between systems of Lambda atoms and
systems of pairwise entangled V atoms. Finally we discuss the
physical differences between our photon cloners and the qubit
cloners considered previously and prove that the bounds on the
fidelity of the clones derived for qubits also apply in our
situation.
\end{abstract}

\section{Introduction}

An ideal quantum cloning machine is a device that produces an
arbitrary number of perfect copies of a given (unknown) quantum
system. Such a device would allow the exact determination of the
quantum state of a system. It has been shown \cite{wzurek} that
such a device would violate the linearity of quantum mechanics
 and also relativistic locality because it would make
superluminal communication possible \cite{gisin,herbert}.

Non-perfect copying, though, can be realized in quantum mechanics.
Since the seminal paper of Bu\v{z}ek and Hillery
\cite{buzekhillery}, quantum cloning has been extensively studied
theoretically. Bruss et al. \cite{brussetal} derived bounds on the
possible fidelity of quantum cloners, Gisin and Massar, and
Bu\v{z}ek and Hillery \cite{gisinmassar} discovered optimal
universal cloning transformations, and finally Werner, and Keyl
and Werner \cite{werner} discussed optimal universal cloning in
great generality.

While optimal cloning was previously discussed in terms of quantum
networks, in a recent paper some of the authors have shown that
optimal universal cloning can be comparatively easily realized via
stimulated emission \cite{simon}. In this scheme the general qubit
to be cloned is represented by the polarization state of a photon.
When cloning is realized via stimulated emission, the fidelity of
the clones is limited by the unavoidable presence of spontaneous
emission. It was shown that the bounds on the fidelity given by
the above-mentioned fundamental principles can nevertheless be
saturated.

In Sec. \ref{lambdas} of the present paper we present a scheme for
{\it optimal} universal cloning based on stimulated emission in
three-level systems of the Lambda type. Our main analytic tool is
the formal equivalence between systems of Lambda atoms and coupled
harmonic oscillators, which is established in Sec. \ref{osc}. In
Sec. \ref{general}, this equivalence is used to analyze the
structure of the transformations realized in detail and to prove
their optimality. More specifically, we explicitly demonstrate
their equivalence to the optimal cloning transformations for
qubits discovered before \cite{buzekhillery,gisinmassar}. In
particular, it will become clear that the atomic states play the
double role of photon source and ancilla, and that the universal
NOT operation is realized in the ancilla states. In the same way,
we show that the down-conversion cloner presented in \cite{simon}
is obtained from the present schemes as a limiting case. In Sec.
\ref{vs} we demonstrate that optimal cloning can also be achieved
with pairwise entangled V atoms, using an interesting equivalence
between the two systems. In Sec. \ref{comparison} we discuss the
physical differences that exist between our stimulated emission
cloners and the qubit cloners considered previously, and we give
an explicit proof that the bounds derived for qubit cloning indeed
apply to our situation as well. Section \ref{conclusions} gives
our conclusions.

\section{Cloning via stimulated emission in Lambda-atoms}
\label{lambdas}

The general principles of universal cloning via stimulated
emission are the following. Consider an inverted medium that can
spontaneously emit photons of any polarization with the same
probability. If a photon (or several) of a given polarization
interacts with such a medium, it stimulates the emission of
photons of the same polarization. In the final photonic state
there will be a majority of photons polarized parallel to the
incoming photon, while some photons will be in the orthogonal
polarization due to spontaneous emission. In this way the photons
in the final state can be considered as clones of the original
incoming photon, where the fidelity of the clones is given by the
relative frequency of photons of the correct polarization in the
final state.

The inverted medium that we will use as a cloning device consists
of an ensemble of Lambda-atoms. These are three-level systems that
have two degenerate ground states $|g_1\rangle$ and $|g_2\rangle$
and an excited level $|e\rangle$. The ground states are coupled to
the excited state by two
 modes of the electromagnetic field, $a_1$ and $a_2$, respectively. These two modes define the
  Hilbert space of our qubit to be cloned, i.e. we want to clone general superposition
   states $(\alpha a_1^\dagger + \beta a_2^\dagger)|0,0\rangle=\alpha|1,0\rangle+\beta |0,1\rangle$.
   We can think of $a_1$ and $a_2$ as being orthogonal polarizations of one photon
   with a specific frequency, but we do not have to restrict ourselves to such a specific example,
   in fact we can think about other systems and other degrees of freedom,
   as long as they are described by the same formalism, e.g. $a_1$ and $a_2$ could also refer to
   the center-of-mass motion (phonons) in an ion trap. In the interaction
   picture, after the usual dipole and
rotating wave approximations,
   the interaction Hamiltonian between
   field and atoms has the following form:
\begin{equation}
\label{Ham1} {\cal H}_i=\gamma \left( a_1 \sum_{k=1}^N |e^k\rangle
\langle g^k_1| + a_2 \sum_{k=1}^N |e^k\rangle \langle g^k_2|
\right) + h.c.=\gamma \left( a_1 \sum_{k=1}^N \sigma^k_{+,1} + a_2
\sum_{k=1}^N \sigma^k_{+,2} \right) + h.c.
\end{equation}
The index $k$ refers to the $k$-th atom. Note that in (\ref{Ham1})
the atoms couple to only one single spatial mode of the
electromagnetic field. In particular this means that spontaneous
emission into all other modes is neglected. Situations where this
is a good approximation can now be achieved in cavity QED
\cite{cavity}. We also assume that the coupling constant $\gamma$
is the same for all atoms, which in a cavity QED setting means
that they have to be in equivalent positions relative to the
cavity mode. Trapping of atoms inside a cavity has recently been
achieved \cite{kimble}. Finally note that our Hamiltonian has no
spatial dependence, which means that the effect of the field on
the motion of the atoms is neglected, their spatial wavefunction
is assumed to be unchanged \cite{bec}.

The Hamiltonian (\ref{Ham1}) is invariant under simultaneous
unitary transformations of the vectors $(a_1,a_2)$ and
$(|g_1\rangle, |g_2\rangle)$ with the same matrix $U$. If one
furthermore chooses an initial state of the atoms that has the
same invariance, then the system behaves equivalently for all
incoming photon polarizations, i.e. universal cloning is achieved.
This can be seen in the following way. Consider an incident photon
in a general superposition state $(\alpha a^\dagger_1 + \beta
a^\dagger_2) |0,0\rangle$. Together with the orthogonal one-photon
state this defines a new basis in polarization space, which is
connected to the original one by a unitary transformation. If the
atomic states are now rewritten in the basis  that is connected to
the original one by the same unitary transformation, then under
the above assumptions the interaction Hamiltonian and initial
state of the atoms look exactly the same as in the original basis.
The initial state where all atoms are excited to $|e\rangle$ has
the required invariance: it is completely unaffected by the
above-mentioned transformations.

We can therefore, without loss of generality, restrict ourselves
to the cloning of photons in mode $a_1$. We consider an initial
state
\begin{equation}
|\Psi_{in}\rangle=\otimes_{k=1}^N |e^k\rangle
\frac{(a_1^\dagger)^m}{\sqrt{m!}}|0,0\rangle, \label{psii}
\end{equation}
i.e. we are starting with $m$ photons of a given polarization, and
we want to produce a certain (larger) number $n$ of clones.

\subsection{The simplest case}

For illustrative purposes let us first consider the simplest case
of one Lambda-atom and one photon polarized in direction $1$:
\begin{equation}
 |\Psi_{in}\rangle=|e\rangle
 a^\dagger_1|0,0\rangle=|e\rangle |1,0\rangle
 =: |{\cal F}_0\rangle
\end{equation}
 To
study the time development, we expand the evolution operator
$e^{-i{\cal H}t}$ into a Taylor series and determine the action of
powers of ${\cal H}$ on the state $|\Psi_{in}\rangle$.
\begin{eqnarray}
{\cal H} |\Psi_{in}\rangle&=&\gamma (|g_1\rangle a_1^\dagger
 |1,0\rangle+|g_2\rangle a_2^\dagger |1,0\rangle)=  \gamma \sqrt{3} \frac{  (\sqrt{2}|g_1\rangle |2,0\rangle + |g_2\rangle
 |1,1\rangle)}{\sqrt{3}}\nonumber  =: \gamma \sqrt{3}
 |{\cal F}_1\rangle \nonumber \\
{\cal H}^2 |\Psi_{in}\rangle&=&\gamma^2 (|e \rangle a_1 \sqrt{2}
 |2,0\rangle + |e\rangle a_2 |1,1\rangle)=3 \gamma^2 |e\rangle
 |1,0\rangle=3 \gamma^2  |{\cal F}_0\rangle
\nonumber \\
 & \ldots & \nonumber \\
 \label{hpowers}
 \end{eqnarray}

 The result is
 \begin{eqnarray}
 \label{onelam}
e^{-i{\cal H} t }|\Psi_{in}\rangle &=& \cos(\gamma \sqrt{3}t
)|e\rangle |1,0\rangle - i \sin{(\gamma \sqrt{3}t)}
(\sqrt{\frac{2}{3}}|g_1\rangle |2,0\rangle +\sqrt{\frac{1}{3}}
 |g_2\rangle |1,1\rangle)\nonumber\\
&=& \cos(\gamma \sqrt{3}t) |{\cal F}_0\rangle -i \sin{(\gamma
 \sqrt{3}t)} |{\cal F}_1\rangle
\end{eqnarray}
$|{\cal F}_0\rangle$ and $|{\cal F}_1\rangle$ denote the states of
the system atom-photons that lie in the subspace with $1$ and $2$
photons respectively.
 $|{\cal F}_0\rangle$ is in the subspace where no cloning has taken place and $|{\cal F}_1\rangle$
 in the one where
one additional photon has been emitted, so that the two photons
can now be viewed as clones with a certain fidelity. This way of
labeling the states will turn out to be convenient below. The
probability that the system acts as a cloner is
$p(1)=\sin^2(\gamma \sqrt{3}t)$. The fidelity $F_1$ of the cloning
procedure can be defined as the relative frequency of photons in
the correct polarization mode in the final state $|{\cal
F}_1\rangle$ (cf. Sec. \ref{comparison}). One finds
\begin{equation}
F_1=\frac{2}{3}\cdot1+\frac{1}{3}\cdot\frac{1}{2}=\frac{5}{6},
\end{equation}
which is exactly the optimal fidelity for a 1-to-2 cloner
\cite{buzekhillery,brussetal}. Actually, the state
\begin{equation}
|{\cal
F}_1\rangle=\sqrt{\frac{2}{3}}|2,0\rangle|g_1\rangle+\sqrt{\frac{1}{3}}|1,1\rangle|g_2\rangle
\label{f1}
\end{equation}
is exactly equivalent to the three-qubit state
\begin{equation}
\sqrt{\frac{2}{3}}|11\rangle|\downarrow\rangle +
\sqrt{\frac{1}{3}}\left(\frac{1}{\sqrt{2}}(|01\rangle+|10\rangle)\right)|\uparrow\rangle
\label{bh}
\end{equation}
produced by the Bu\v{z}ek-Hillery cloner, see \cite{buzekhillery},
Eq. (3.29b). The equivalence is established, if the photonic
states in Eq. (\ref{f1}) are identified with the corresponding
{\it symmetrized} two-qubit states (both photons in mode 1 means
both qubits in state $|1\rangle$, one photon in each mode means
one qubit in state $|1\rangle$, one in state $|0\rangle$) in Eq.
(\ref{bh}), while the atomic states $|g_1\rangle$ and
$|g_2\rangle$ are identified with the states $|\downarrow\rangle$
and $|\uparrow\rangle$ of the ancillary qubit. This is another way
of proving the optimality of Eq. (\ref{f1}). Note that in our case
the {\it universality} follows directly from the symmetry of
initial state and Hamiltonian, as explained above. In the
following we show that a similar equivalence holds between our
cloning scheme and the Gisin-Massar cloners in the completely
general case (arbitrary numbers of photons and atoms).

\subsection{Equivalence to coupled harmonic oscillators}

\label{osc}

We now turn to the discussion of the general case, i.e. we
consider the initial state (\ref{psii}). We are going to show the
equivalence of our system defined by (\ref{Ham1}) and (\ref{psii})
to a system of coupled harmonic oscillators. First note that both
the initial state (\ref{psii}) and the Hamiltonian (\ref{Ham1})
are invariant under permutations of the atoms, which implies that
the state vector of the system will always be completely
symmetric. Furthermore the Hamiltonian (\ref{Ham1}) can be
rewritten as
\begin{equation}
{\cal H}=\gamma \left( a_1 J_{+,1} + a_2 J_{+,2} \right) + h.c.
\end{equation}
in terms of ``total angular momentum'' operators
\begin{equation}
J_{+,r}=  \sum_{k=1}^N \sigma^k_{+,r} =     \sum_{k=1}^N
|e^k\rangle \langle g^k_r|           \hspace{1cm}  (r=1,2),
\end{equation}
By the above considerations one is led to use a Schwinger type
representation \cite{schwinger} for the angular momentum
operators:
\begin{equation}
J_{+,r}= b_r c^{\dagger}    \hspace{1cm} (r=1,2),
\label{schwinger}
\end{equation}
where $c^{\dagger}$ is a harmonic oscillator operator creating
``$e$'' type excitations, while $b_1$ destroys ``$g_1$''
excitations. Note that $J_{+,1}$ and $J_{+,2}$ share the operator
$c^{\dagger}$ because both ground levels $g_1$ and $g_2$ are
connected to the same upper level $e$ by the Hamiltonian
(\ref{Ham1}), and correspondingly for the Hermitian conjugates. In
terms of these operators, (\ref{Ham1}) acquires the form
\begin{equation}
{\cal H}_{osc}=\gamma ( a_1 b_1 +  a_2 b_2)c^{\dagger}+ h.c.,
\label{Hampd1}
\end{equation}
while the initial state (\ref{psii}) is now given by
\begin{equation}
|\psi_i\rangle=\frac{(a_1^{\dagger})^m}{\sqrt{m!}}\frac{(c^{\dagger})^N}{\sqrt{N!}}|0\rangle=
|m_{a1},0_{a2},0_{b1},0_{b2},N_c\rangle\equiv|m,0,0,0,N\rangle.
\label{psiinew}
\end{equation}
Actually, for reasons that will become apparent below, it is
slightly more convenient for our purposes to use the following
Hamiltonian instead of (\ref{Hampd1}):
\begin{equation}
{\cal H}=\gamma (a_1 b_2 -  a_2 b_1) c^{\dagger}+ h.c.,
\label{Hampd}
\end{equation}
which can be obtained from (\ref{Hampd1}) by a simple unitary
transformation in mode $b$, corresponding to a simple redefinition
of the atomic states in (\ref{Ham1}). This is the Hamiltonian that
is going to be used in the rest of this paper. The invariance
properties of (\ref{Hampd}) are linked to those of (\ref{Ham1}) or
equivalently (\ref{Hampd1}) discussed above: (\ref{Hampd}) is
invariant under simultaneous identical SU(2) transformations in
modes $a$ and $b$ (because the determinant of such a
transformation is equal to unity), while a phase transformation in
either mode can be absorbed into $\gamma$. This ensures the
universality of the cloning procedure.

We are now dealing with five harmonic oscillator modes defined by
the operators $c, b_1, b_2, a_1$, and $a_2$. Action of
(\ref{Hampd}) on (\ref{psiinew}) generates Fock basis states of
the general form
\begin{equation}
|(m+j)_{a1},i_{a2},i_{b1},j_{b2},(N-i-j)_c\rangle=|m+j,i\rangle_{photons}|i,j,N-i-j\rangle_{atoms}.
\end{equation}
Remember that $a_1$ is now coupled to $b_2$ etc. Expressed in
terms of individual atoms, $|i,j,N-i-j\rangle_{atoms}$ is the
completely symmetrized state with  $i$ atoms in level $g_1$, $j$
atoms in level $g_2$, and $N-i-j$ atoms in level $e$. The
correctness of (\ref{schwinger}) can be checked by explicit
application of left hand side and right hand side to such a
general state, written in terms of the individual atoms and in
terms of harmonic oscillator eigenstates respectively (see
Appendix A).

 Note that the use of the Schwinger
representation is only convenient because the initial state of the
atomic system in (\ref{psii}) is completely symmetric under
permutation of the atoms.

Studying the Hamiltonian in the form (\ref{Hampd}) instead of
(\ref{Ham1}) is helpful in several respects. The number of atoms
$N$ that is explicit in the Hamiltonian (\ref{Ham1}) now appears
only as a part of the initial conditions of our system, which
makes it easy to treat the general case of N atoms in one go. We
will do this in the next subsection.

Furthermore, the connection to cloning by parametric
down-conversion (PDC) as proposed in \cite{simon} is now obvious.
The Hamiltonian (\ref{Hampd}) can also be seen as a Hamiltonian
for down-conversion with a quantized pump-mode described by the
operator $c$, while $a_r$ and $b_r$ are the signal and idler modes
respectively, where $r$ labels the polarization degree of freedom.
There is only one difference between (\ref{Hampd}) and the
Hamiltonian used in \cite{simon} (see Eq. 6 of that reference): in
\cite{simon} the operator $c$ of (\ref{Hampd}) is replaced by a
c-number. In the context of down-conversion, this corresponds to
the limit of a classical pump field. Thus the PDC scheme, which
was shown to achieve optimal universal cloning in \cite{simon}, is
obtained as a limiting case from the schemes discussed here.

In passing we note that the above dynamical equivalence generalizes
to atoms with more than $2$ ground-states $|g_n\rangle$ that are
coupled each to a different degree of freedom of photons $a_n$. By
similar arguments a system of $N$ identical atoms with $r$ ground
states $\{|g_1\rangle,\ldots,|g_r\rangle\}$ governed by a
Hamiltonian
\begin{equation}
{\cal H}^r=\gamma \sum_{k=1}^N \sum_{n=1}^r |e^k\rangle \langle
g_n^k| a_n +h.c.
\end{equation}
is equivalent to a system of $r+1$ coupled harmonic oscillators with
lowering operators $c$ and $b_1, \ldots , b_r$ governed by the interaction Hamiltonian
\begin{equation}
{\cal H}^r_{osc}=\gamma \sum_{n=1}^r cb_n^\dagger a_n^\dagger +
h.c.
\end{equation}

\subsection{Cloning of $m$ photons with $N$ Lambda-atoms: Proof of optimality}

\label{general}

We are now going to show that the system defined by
(\ref{psiinew}) and (\ref{Hampd}) indeed realizes optimal cloning
for arbitrary $N$ and $m$. The idea of the proof is the following.
After evolution in time the system that started with a certain
photon number $m$ will be in a superposition of states with
different total photon numbers, where total means counting photons
in mode $a_1$ and $a_2$, i.e. both ``good'' and ``bad'' copies. We
will show that the general form of the state vector after a time
interval $t$ is
\begin{equation}
|\Psi(t)\rangle = e^{-i {\cal H} t}
 |\Psi_{in}\rangle =\sum_{l=0}^N f_l(t) |{\cal F}_l\rangle,
\end{equation}
where $l$ denotes the number of {\it additional} photons that have
been emitted and
\begin{equation}
\label{Fs} |{\cal F}_l\rangle:={{m+l+1}\choose {l}}^{-\frac{1}{2}}
\sum_{i=0}^l (-1)^i \sqrt{{m+l-i} \choose {m}}
 |(m+l-i)_{a1},i_{a2},i_{b1},(l-i)_{b2},(N-l)_c\rangle.
\end{equation}
Note that the number of photons can never become smaller than $m$
since all the atoms start out in the excited state. $|{\cal F}
_l\rangle$ is a normalized state of the system with $m+l$ photons
in total. To see that $|{\cal F}_l\rangle$ is properly normalized
note that $\sum_{i=0}^l {{m+i} \choose {m}} = {{m+l+1} \choose
{l}}$.

The states $|{\cal F}_l\rangle$ are formally identical to the
states obtained in \cite{unot}, which have been shown to realize
optimal universal cloning and the optimal universal NOT
simultaneously. The ideal universal NOT is an operation that
produces the orthogonal complement of an arbitrary qubit. Like
perfect cloning, it is prohibited by quantum mechanics. The
transformation in \cite{unot} links universal cloning and
universal NOT (anti-cloning): the ancilla qubits of the cloning
transformation are the anti-clones. In our case, the clones are
the photons in the $a$-modes and the anti-clones are the atoms in
the $b$-modes (atomic ground states $g_1$ and $g_2$). From the
Hamiltonian (\ref{Hampd}) and (\ref{Fs}) it is clear that for
every ``good'' emitted photon-clone (in mode $a_1$) there is an
excitation in mode $b_2$ which corresponds to an anti-clone
(atomic ground state $|g_2\rangle$). The only difference to the
states in \cite{unot} is the presence of the fifth harmonic
oscillator mode $c$, describing the ``e'' type excitations, which
counts the total number of clones that have been produced (equal
to the number of atoms having gone to one of the ground states)
and doesn't affect any of the conclusions.

A distinguishing feature of our cloner is that the output state
(\ref{Fs}) is a superposition of states with different total
numbers of clones. Cloning with a certain fixed number of produced
copies can be realized by measuring the number of atoms in the
excited state $|e\rangle$ (corresponding to mode $c$) and
post-selection.

To see that the $|{\cal F}_l\rangle$ are indeed the output of an
optimal cloner, let us calculate the fidelity of the cloning,
given by the mean relative frequency of photons in the correct
mode ($a_1$ in our case).
 In the state
$|m+l-i,i\rangle_{photons}$ the relative frequency of correct
photons is $(m+l-i)/(m+l)$. Therefore
\begin{eqnarray}
F_l &=& {{m+l+1}\choose {l}}^{-1} \sum_{i=0}^l {{m+l-i} \choose
{m}}\frac{m+l-i}{m+l} = \frac{m(m+2) + l(m+1)}{(m+l)(m+2)}
\end{eqnarray}
which corresponds to the fidelity of an optimal universal
$m\rightarrow m+l$ cloner \cite{brussetal}. Note again that the
universality in our case follows from the symmetry of the
Hamiltonian and the initial atomic state.

To prove that the system is indeed always in a superposition of
the states $|{\cal F}_l\rangle$ as in Eq. (\ref{Fs}) we use
induction: The initial state of the system is
$|\Psi_{in}\rangle=|{\cal F}_0\rangle$. Now we will show that if
$|\Phi\rangle$ is a superposition of states $|{\cal F}_l\rangle$
then ${\cal H} |\Phi\rangle$ is so, too. Then, since
$|\Psi(t)\rangle=e^{-i {\cal H} t} |\Psi_{in}\rangle=\sum_{p}
\frac{(-i{\cal H} t)^p}{p!} |\Psi_{in}\rangle$ this implies that
$|\Psi(t)\rangle$ will be a superposition of $|{\cal F}_l\rangle$.
Explicit calculation shows that
\begin{eqnarray}
{\cal H} |{\cal F}_l\rangle & = & \gamma
(\sqrt{(l+1)(N-l)(m+l+2)}|{\cal F}_{l+1}\rangle+
\sqrt{l(N-l+1)(m+l+1)}  |{\cal F}_{l-1}\rangle) \quad 1 \leq l < N
  \nonumber \\{\cal H} |{\cal F}_0\rangle & = & \gamma \sqrt{N(m+2)}|{\cal F}_{1}\rangle \nonumber \\
{\cal H} |{\cal F}_N\rangle & = & \gamma \sqrt{N(m+N+1)} |{\cal
F}_{N-1}\rangle \label{rec}
\end{eqnarray}
which completes the proof.

Note that the form of the coefficients $f_{l}(t)$ didn't play any
role in our proof. Actually, the $f_l$ are in general hard to
determine exactly. Solutions have been found in limiting cases.
For the limit of a classical pump field ($c$ replaced by a
c-number), the solution can be found by standard methods and is
given in \cite{simon}. The solution in the case of large incoming
photon numbers ($m\gg N$) is presented in Appendix B.

Let us pause here for a moment and summarize what we have found.
Our system consisting of an ensemble of Lambda-atoms in the
excited state is indeed equivalent to a superposition of optimal
cloning machines a la Bu\v{z}ek-Hillery or Gisin-Massar, producing
various numbers of clones. The atoms play the double role of
photon source and of ancilla, the atomic ground states can be
identified with the ancilla states in the qubit cloners. As for
the corresponding qubit cloners, those ancillary atoms can also be
seen as the output of a universal NOT gate. On the other hand, the
atoms that end up in the excited state provide information about
the number of clones that has actually been produced. This can be
used to realize cloning with a fixed number of output clones by
post-selection.

\section{The equivalence between pairs of V-atoms and Lambda-atoms}
\label{vs}

In this section we present an alternative (but similar) way of
realizing optimal universal cloning that uses entangled pairs of
V-atoms instead of Lambda atoms. We prove optimality by showing
that the system can be exactly mapped onto the system with Lambda
atoms that we discussed above.

The two degenerate upper levels of each V-atom, $|e_1\rangle$ and
$|e_2\rangle$, are coupled to the ground state $|g\rangle$ via the
two orthogonal
 modes $a_1$ and $a_2$ respectively. The Hamiltonian describing the interaction of atom and field is:
\begin{equation}
\label{Ham2} {\cal H}_V= \gamma \left(a^\dagger_1 \sum_{k=1}^N
|g^k\rangle \langle e^k_1| +a^\dagger_2 \sum_{k=1}^N |g^k\rangle
\langle e^k_2| \right) + h.c.= \gamma \left(a^\dagger_1
\sum_{k=1}^N \sigma^k_{-,1} +a^\dagger_2 \sum_{k=1}^N
\sigma^k_{-,2} \right) + h.c.
\end{equation}
It arises from similar assumptions as (\ref{Ham1}). In contrast to
before we now choose an entangled state of the atoms as the
initial state. This is motivated by the fact that the initial
atomic state has to be a {\it singlet} under polarization
transformations in order for our cloning device to be again
universal.

Let us first examine the simplest case of two entangled V-atoms,
$A$ and $B$, and one incoming photon. The initial state of the
system is
\begin{equation}
|\Psi_{in}\rangle=\frac{1}{\sqrt{2}}(|e_1^A e_2^B\rangle-|e_2^A
 e_1^B\rangle)\otimes |1,0\rangle
\end{equation}
Developing the time evolution operator $e^{-i{\cal H_V}t}$ into a
power series, one finds easily:
\begin{eqnarray}
e^{-i{\cal H_V} t }|\Psi_{in}\rangle &=& \cos(\gamma \sqrt{3}t
 ) \frac{|e_1^A e_2^B\rangle - |e_2^A
e_1^B\rangle}{\sqrt{2}} |1,0\rangle - \nonumber \\ & & i
\sin{(\gamma \sqrt{3}t)} \left(\sqrt{\frac{2}{3}} \frac{|g^A
e_2^B\rangle - |e_2^A g^B\rangle}{\sqrt{2}} |2,0\rangle
+\sqrt{\frac{1}{3}} \frac{|e_1^A g^B\rangle - |g^A
e_1^B\rangle}{\sqrt{2}} |1,1\rangle \right) \label{twov}
\end{eqnarray}
With the substitution
\begin{eqnarray}
\label{subst} \frac{|e_1^A e_2^B\rangle-|e_2^A
e_1^B\rangle}{\sqrt{2}} \longrightarrow |\tilde{e}\rangle
\nonumber\\ \frac{|g^A e_2^B\rangle-|e_2^A g^B\rangle}{\sqrt{2}}
\longrightarrow |\tilde{g_1}\rangle \nonumber\\ \frac{|e_1^A
g^B\rangle-|g^A e_1^B\rangle}{\sqrt{2}} \longrightarrow
|\tilde{g_2}\rangle
\end{eqnarray}
the state (\ref{twov}) has exactly the same form as the
corresponding state (\ref{onelam}) for one Lambda-atom, which
implies that it also implements optimal universal $1\rightarrow 2$
cloning.

Actually, the correspondence goes much further. Consider an
initial atomic state consisting of $N$ pairs of V-atoms, where
each pair is in a singlet state:
\begin{equation}
|\psi_i\rangle=\otimes_{k=1}^N |\tilde
  e^k\rangle
\end{equation}
with $|\tilde e\rangle$ as defined in (\ref{subst}).

It is easy to see that the action of the Hamiltonian (\ref{Ham2})
on each pair only
  generates one of the three antisymmetric atomic states in Eq. (\ref{subst}). Because of the
  invariance of the Hamiltonian under permutations, and in particular under the exchange
  of two atoms belonging to the same pair, transitions between states with different symmetry
  properties are impossible.
  In fact, with the identification (\ref{subst}) the Hamiltonian (\ref{Ham2}) has exactly the same form
  as the
  Hamiltonian for Lambda-atoms (\ref{Ham1}).
The analysis made for Lambda atoms in Section \ref{lambdas} now
goes through unchanged and we obtain the same cloning properties
of a system of pairwise entangled V-atoms as we had before for
Lambda-atoms, i.e. we have found another way of realizing optimal
universal cloning. Although this scheme would without doubt be
more difficult to realize experimentally, we believe that the
underlying equivalence between the two systems is interesting and
may be useful in other contexts as well.

\section{Cloning of photons versus cloning of qubits}

\label{comparison}

In this section we are going to discuss the physical differences
that exist in spite of the formal equivalence proven above between
our photon cloners based on stimulated emission and the qubit
cloners as usually considered \cite{buzekhillery,gisinmassar}. In
particular, we will show that the claim that optimal cloning is
realized by our devices is justified in spite of these
differences.

In most of the previous work cloning was discussed in terms of
quantum networks. In general, the situation considered in these
papers is the following: one has a certain number of qubits that
are localized in different positions, which makes them perfectly
distinguishable. At the beginning, some of those qubits are the
systems to be cloned, the others play the role of ancillas. After
the cloning procedure, which consists of several joint operations
on the qubits that can be expressed in terms of quantum gates,
some of the qubits are the clones, the rest are ancillas, which
for a specific form of the optimal cloning transformation can also
be seen as outputs of the universal NOT operation. As a
consequence of localization, it is possible to address individual
clones.

In our stimulated emission cloners, the situation is different.
All input systems (photons) are in the same spatial mode (called
mode $a$ in this paper), and, even more importantly, all clones
are produced into that mode. Note that this is completely
unavoidable if stimulated emission is to be used. One can say that
this is the price one has to pay for the great conceptual
simplicity of the cloning procedure itself.

However, having all clones in the same spatial mode is not
necessarily an important disadvantage. For example, if perfect
cloning of that kind were possible, one could still determine the
polarization of the original photon to arbitrary precision by
performing measurements on the clones. This would still make
superluminal communication possible \cite{start}. If one wants to
distribute the clones to different locations, this can for example
be achieved using an array of beam splitters. However, this does
not lead to a situation where one can be sure to have exactly one
photon in each mode. If one wants to have at most one photon in
each mode, the array has to have many more output modes than there
are photons.

Another distinguishing feature of our cloners compared to the
usual qubit cloners is the fact that the same procedure is used to
produce different numbers of clones. While in the qubit case the
network to be used depends on the number of desired clones, in our
case the final state is a superposition of states with different
numbers of clones. Of course, the average number of clones
produced depends on the number of atoms present in the system and
the interaction time. As discussed in Sec. \ref{lambdas} cloning
with a fixed number of output clones can be achieved by
post-selection based on a measurement of the number of excited
atoms in the final state.

The formal equivalence between the qubit cloners and our one-mode
cloners can arise because the output state produced by the optimal
qubit cloners is completely symmetric under the exchange of clones
\cite{buzekhillery,gisinmassar}. Because of the bosonic nature of
the photons there is a one-to-one-mapping between completely
symmetric qubit states and photonic states. For a completely
symmetric qubit state the two concepts of relative frequency of
qubits in the ``correct'' basis state and of single-particle
fidelity are equivalent. This can be seen in the following way.
Let $|\psi\rangle$ denote the state that is to be copied. Then the
usual definition of the (single-particle) cloning fidelity is
\begin{equation}
F=\langle\psi|\rho_{red} |\psi\rangle,
\end{equation}
where $\rho_{red}$ is the reduced density matrix of one of the
clones, say the first one, i.e.
\begin{equation}
\rho_{red}=\mbox{Tr}_{2,3,...,N}\left[\, \rho \, \right]
\end{equation}
 Then $F$ can also be expressed as
\begin{equation}
F=\mbox{Tr}\left[\, \rho \; |\psi\rangle\langle\psi|_1\otimes
I_2\otimes...\otimes I_N\right]. \label{F}
\end{equation}
On the other hand, the relative frequency of qubits in the state
$|\psi\rangle$ can be written as
\begin{equation}
\frac{1}{N}\mbox{Tr}\left[\, \rho \,
\left(|\psi\rangle\langle\psi|_1\otimes I_2\otimes...\otimes I_N +
I_1 \otimes |\psi\rangle\langle\psi|_2 \otimes ... \otimes I_N +
... + I_1 \otimes ... \otimes |\psi\rangle\langle\psi|_N \right)
\right]. \label{rel}
\end{equation}
If $\rho$ is invariant under exchange of any two clones, it is
obvious that (\ref{rel}) is equal to (\ref{F}), i.e. for symmetric
cloners the two concepts are completely equivalent. This justifies
our definition of fidelity via the relative frequency in the case
of photon cloning (cf. Sec. \ref{lambdas}).

Let us finally address the issue of optimality in the context of
stimulated emission cloners. In this paper we have shown the
formal equivalence of our scheme and the optimal schemes for qubit
cloning. As a consequence, the fidelity of the clones saturates
the bounds derived for the cloning of qubits. However, it is not
entirely obvious that the bounds derived for the situation of
distinct well-localized qubits also apply to our situation. Could
one maybe achieve even higher fidelity in our one-mode case? The
following argument shows that the bounds indeed apply in our
situation as well, i.e. that photon cloning is not allowed to be
better than qubit cloning.

Let us assume that we had a single-mode cloning machine that
clones photons with a better fidelity than given by the bounds for
qubits. Consequently, the relative frequency of ``correct''
photons has to exceed the bound for at least one value of the
final total photon number $M$. This is obvious if $M$ has been
fixed by post-selection. Otherwise the fidelity has to be defined
as the average of the relative frequencies over all final total
photon numbers. This average can only exceed the bound for qubits
if the bound is violated for at least one particular value $M$ of
the final photon number.

As a consequence, we have a universal map from the $N$-photon
Hilbert space to the $M$-photon Hilbert space that achieves a
relative frequency of correct photons in the final state that is
higher than the qubit bound. But the existence of such a map is
equivalent to the existence of a universal map from the totally
symmetric $N$-qubit space to the totally symmetric $M$-qubit space
with a single-particle fidelity equal to the relative frequency.
The existence of the latter map is excluded by the theorems on
cloning of qubits \cite{werner}. This justifies our claim that the
schemes presented in the previous sections realize {\it optimal}
cloning of photons.

\section{Conclusions and Outlook}
\label{conclusions}

In this paper we have shown that optimal universal cloning can be
realized via stimulated emission in three-level systems. The
permutation symmetry of the interaction allowed to map our system
onto bosonic modes independent of the number of atoms used.
Furthermore, we have found an equivalence between single Lambda
atoms and entangled pairs of V systems, which might be fruitful in
other contexts as well.

The connection between stimulated emission and optimal cloning is
remarkable. Our results show that a task previously discussed in
terms of rather complicated quantum networks can be realized in an
elegant way using basic quantum systems and interactions. While it
was clear from the beginning that perfect cloning is prohibited by
fundamental principles, it is interesting to see how this
impossibility arises in a concrete physical system. In our case,
the physical process limiting the fidelity of the clones is
spontaneous emission. It is fascinating that in this way
spontaneous emission ensures that there cannot be any superluminal
communication.

 It might be interesting to investigate
possible experimental realizations of our proposal, e.g. using a
combination of cavity QED and Bose-Einstein condensates. This
would potentially allow the creation of macroscopic numbers of
clones.

Quantum cloners are often discussed in the context of
eavesdropping in quantum cryptography. Currently all cryptography
schemes rely on photons. Therefore devices based on the principles
presented here could be useful to a future eavesdropper.

We would like to thank  A. Zeilinger for stimulating discussions
on various aspects of the present work, V. Bu\v{z}ek for pointing
out that what we are doing can be seen as use of the Schwinger
representation, and A. Karlsson for motivating us to discuss the
physical peculiarities of our photon cloners. Furthermore we thank
D. Bacon, \v{C}. Brukner, I. Cirac, and S. Stenholm for useful
discussions. This work was supported by the Austrian Science
Foundation (FWF), project no. S6504 and F1506, by the U.S. ARO
under DAAG55-98-1-0371 and by U.S. NSF DMS-9971169.

\appendix
\section{Schwinger representation}

As noted above, the action of the Hamiltonian (\ref{Ham1}) on the
initial state (\ref{psii}) only generates completely symmetric
states of the atomic system. These states have the general form
\begin{equation}
\label{atomic}  {{N} \choose {i,j}}^{-1/2} \sum_{\alpha}
|g_1^{\alpha_1},g_1^{\alpha_2},\ldots,g_1^{\alpha_i},g_2^{\alpha_{i+1}},\ldots,g_2^{\alpha_{i+j}},
e^{\alpha_{i+j+1}},\ldots,e^{\alpha_N}\rangle =:| i, j, N-i-j
\rangle_{atoms}
\end{equation}
where the sum is over all arrangements $\alpha$ of the $N-i-j$
levels $|e\rangle$, the $i$ levels $|g_1\rangle$, and the $j$
levels $|g_2\rangle$ on the $N$ atoms, and ${{N} \choose
{i,j}}=\frac{N!}{i!j!(N-i-j)!}$ is the multinomial coefficient
giving the number of such arrangements.

Now study the action of a typical term in the Hamiltonian
(\ref{Ham1}) on the system whose state we will write as\\
$|i,j,N-i-j\rangle_{atoms} \otimes |m+i,j\rangle_{photons}$:
\begin{eqnarray}
& &(\sum_{k=1}^N |g_1^k\rangle \langle e^k|) a_1^\dagger
|i,j,N-i-j\rangle_{atoms} \otimes |m+i,j\rangle_{photons}
\nonumber
\\ &=& \sum_{k=1}^N |g^k_1\rangle \langle e^k|
\sqrt{\frac{i!j!(N-i-j)!}{N!}} \sum_\alpha
|g_1^{\alpha_1},\ldots,g_1^{\alpha_i},g_2^{\alpha_{i+1}},\ldots,e^{\alpha_N}\rangle
\otimes a_1^\dagger |m+i,j\rangle_{field} \nonumber \\ &=& (i+1)
\sqrt{\frac{i!j!(N-i-j)!}{N!}} \sum_\alpha
|g_1^{\alpha_1},\ldots,g_1^{\alpha_i},g_1^{\alpha_{i+1}},g_2^{\alpha_{i+2}},\ldots,e^{\alpha_N}\rangle
\otimes a_1^\dagger |m+i,j\rangle_{field} \nonumber \\ &=&
\sqrt{i+1} \sqrt{N-i-j}
\sqrt{\frac{(i+1)!j!(N-i-j-1)!}{N!}}\sum_\alpha
|g_1^{\alpha_1},\ldots,g_1^{\alpha_i},g_1^{\alpha_{i+1}},g_2^{\alpha_{i+2}},\ldots,e^{\alpha_N}\rangle
\otimes a_1^\dagger |m+i,j\rangle_{field} \nonumber \\ &=&
\sqrt{i+1} \sqrt{N-i-j} |i+1,j,N-i-j-1\rangle_{atoms} \otimes
a_1^\dagger |m+i,j\rangle_{photons}
\end{eqnarray}
Here the factor $(i+1)$ arises from the number of different
configurations that a given arrangement $\alpha$ can be reached
by. This shows that this term acts exactly like a term
$a_1^\dagger b_1^\dagger c$. Similar calculations can be made for
the other terms in the Hamiltonian. Together, they justify the
Schwinger representation (\ref{schwinger}).

\section{Limit of large photon number}

Here we determine the coefficients $f_{l}(t)$ of Eq. (\ref{inFs})
in the limit of large $m$ ($m>>N$, many incoming photons, small
number of atoms). For that case, the recursion (\ref{rec}) becomes
\begin{eqnarray}
{\cal H} |{\cal F}_l\rangle & = & \gamma
\sqrt{m}(\sqrt{(l+1)(N-l)}|{\cal
  F}_{l+1}\rangle+  \sqrt{l(N-l+1)}  |{\cal F}_{l-1}\rangle) \quad 1 \leq
  l < N
  \nonumber \\{\cal H} |{\cal F}_0\rangle & = & \gamma \sqrt{m}\sqrt{N}|{\cal
  F}_{1}\rangle \nonumber \\
{\cal H} |{\cal F}_N\rangle & = & \gamma \sqrt{m}\sqrt{N} |{\cal
F}_{N-1}\rangle \label{recm}
\end{eqnarray}
It is possible to diagonalize the ``transfer'' matrix $A$ acting
on
 the vector $(f_0,\ldots,f_N)$
 that corresponds to the action of ${\cal H}$ on
 $|\Psi\rangle=\sum_{l=0}^N f_l |{\cal F}_l \rangle$:
 $A_{l,l+1}=\gamma \sqrt{m} \sqrt{(l+1)(N-l)} =A_{l+1,l}$. This allows
 to exponentiate $A$ and to determine the final state of the system
 after a time $t$:
\begin{equation}
\label{soln} |\Psi(t)\rangle=\sum_{l=0}^N (-i)^l \sqrt{{{N}
\choose {l}}} \cos^{N-l}(
 \gamma \sqrt{m} t) \sin^l( \gamma \sqrt{m} t) |{\cal F}_l\rangle
\end{equation}
Differentiating (\ref{soln}) and using (\ref{recm}) one can show
that this state fulfills Schroedinger's equation with the correct
initial condition.

In this big-$m$-limit the probability to observe the system as an
$m\rightarrow m+l$ cloner (i.e. the probability that $l$
additional photons are emitted) is
\begin{equation}
p(l)={{N} \choose {l}} \cos^{2(N-l)}(\gamma \sqrt{m} t) \sin^{2l}(
\gamma \sqrt{m} t)
\end{equation}
This is a binomial distribution with a probability $\sin^2 (\gamma
\sqrt{m} t)$ for each atom to emit a photon. Setting $N=1$ or
comparison with Eq. (\ref{onelam}) shows that this is identical to
the probability for the case of only one atom in the case of large
$m$. This means that in this limit each atom interacts
independently with the electromagnetic field, because the effect
of the other atoms on the field is negligible. In the short-time
limit $p(l)=O(t^{2l})$. Furthermore the expected average number of
``clones'' $N_c=\sum_{l=0}^N l p(l)=N \sin^2(\gamma \sqrt{m} t)$
oscillates with an $m$-dependent frequency.



\begin{references}
\bibitem{bec}  This leads to the question what the
spatial wavefunction could be. The most fascinating possibility is
probably to imagine a BEC inside a cavity.
\bibitem{start} It may be
interesting to note that in the paper that started the whole
discussion about quantum cloning, Herbert \cite{herbert}
considered cloning via stimulated emission and therefore
necessarily into a single spatial mode.
\end{references}
\end{document}